# Data-driven Inverter-based Volt/VAr Control for Partially Observable Distribution Networks

Tong Xu, *Student Member, IEEE*, Wenchuan Wu, *Fellow, IEEE*, Yiwen Hong, Junjie Yu, Fazhong Zhang

***Abstract*—For active distribution networks (ADNs) integrated with massive inverter-based energy resources, it is impractical to maintain the accurate model and deploy measurements at all nodes due to the large-scale of ADNs. Thus, current models of ADNs are usually involving significant errors or even unknown. Moreover, ADNs are usually partially observable since only a few measurements are available at pilot nodes or nodes with significant users. To provide a practical Volt/Var control (VVC) strategy for such networks, a data-driven VVC method is proposed in this paper. Firstly, the system response policy, approximating the relationship between the control variables and states of monitoring nodes, is estimated by a recursive regression closed-form solution. Then, based on real-time measurements and the newly updated system response policy, a VVC strategy with convergence guarantee is realized. Since the recursive regression solution is embedded in the control stage, a data-driven closed-loop VVC framework is established. The effectiveness of the proposed method is validated in an unbalanced distribution system considering nonlinear loads where not only the rapid and self-adaptive voltage regulation is realized but also system-wide optimization is achieved.**

## I. Introduction

Distributed energy resources (DERs) such as rooftop photovoltaic (PV), residential wind turbines, energy storage systems, have brought considerable benefits to the society reducing the growing gap between energy demand and limited fossil fuels in the past decades. Nonetheless, the increasing penetration of DERs will cause significant and unexpected voltage violation to the modern ADNs due to the uncertainty of renewable energy generation or load fluctuation [1]. Specifically, the reverse power flow introduced by excessive PV injection could result in voltage violations [2] and challenges the operation of the grid [3], [4].

The typical solution is VVC which is essentially a special case of optimal power flow (OPF) with fixed active power and optimal portfolio of reactive power to optimize voltage profile and decrease network losses. Previously, there are abundant researches on optimizing legacy voltage regulating devices such as on-load tap-changer transformers, voltage regulators and capacitor banks [5]. However, legacy devices operate at the slower time-scale (at most minute-level) and are insufficient to provide instant support for the abrupt voltage violation issue. Therefore, smart inverters are getting more attention recently. Smart inverters are equipped with flexible and exceptionally fast (second-level) voltage support functionalities with spare capacity of reactive power [6]. As a matter of fact, the requirement of the smart inverter voltage support is included in new IEEE 1547-2018 standard as well [7]. Since DERs are mostly electronic-interfaced with the power system thus there is great potential for smart inverters to participate into online VVC.

To resolve the non-convexity of power flow, several invert-based VVC methods are developed as suitable solutions for online implementation. A distributed inverter control strategy is presented in [8] to realize the global optimum in unbalanced distribution networks based on linear approximations of power flow. [9] reports a semi-definite programming (SDP) strategy to relax the non-convexity of OPF with global convergence for inverter control. In [10], both nonlinearity of power flow and loads (ZIP model) are relaxed to provide optimal and feasible solution for VVC. Real-time optimization methods for VVC are also presented to cope with the presence of fluctuating loads and volatile renewables [11]. For purely decentralized VVC control which is local, the inverse optimal droop-like method is fairly robust to modeling assumptions and minimizes a combination of power losses and voltage deviations [12]. Reference [13] provides a better transient performance than droop-like method via projected subgradient algorithm. For distributed VVC control which requires coordination, the approach by means of regularization and a distributed dual ascent method is reported in [14] to maximize the distance to voltage collapse. A distributed dual ascent method is presented as well with guaranteed convergence to the operating region where both reactive power limits and voltage constraints are satisfied [15].

Most previous researches assume the complete awareness of the system model where parameters of branches, nonlinear loads, etc. are assumed as prior knowledge. In conventional VVC, global optimum based on the ideal model is solved first then the results are transmitted to local controllers as reference signal to realize the voltage regulation in physical grid [16]. Nevertheless, in real distribution networks, those parameters are probably inaccurate or even unknown instead. Moreover,

This work was supported by the China Southern Power Grid Corp. Research Project (The demonstration and application of the virtual power plant intelligent operation and management platform with source-grid coordination, No. GDKJXM20185069(032000KK52180069)). (*Corresponding Author: Wenchuan Wu*)

T. Xu is with Southwest Electric Power Design Institute Co., Ltd. Of China Power Engineering Consulting Group, Chengdu, Sichuan, 610021, China and Tsinghua University, Beijing, 100084, China (e-mail: taterx@foxmail.com).

W. Wu is with the Department of Electrical Engineering, State Key Laboratory of Power Systems, Tsinghua University, Beijing, 100084, China (e-mail: wuwench@tsinghua.edu.cn).

Y. Hong, J. Yu and F. Zhang are with Guangdong Power Grid Zhongshan Power Supply Bureau, Zhongshan, Guangdong, 528400, China.

for a certain ADN, the model of its host transmission network is unable to be acquired as well. Thus, it is necessary to develop novel VVC approaches which do not depend on accurate network models.

Recently, real-time measurements are introduced in the optimization procedure to provide more reasonable solutions. Measurements are used to replace voltage magnitudes which traditionally should be computed based on nonlinear power flow equations in OPF related optimizations to realize voltage regulation [17], power balance[18] and frequency recovery [19]. Results of single step in the iterative solution will be applied to the physical system as commands. The systematic real-time optimization theory is proposed in [20]-[22] to provide convergence analysis for time-varying OPF solutions. A more detailed framework considering various aggregations of DERs and three-phase operated ADNs is demonstrated in [23]. In [24], the real-time feedback optimization controller is found to be robust with respect to model mismatch via experimental validation. Although real-time optimization methods provide rapid update for online OPF which can be viewed as the partially data-driven method, the gradients in those algorithms are still estimated by ideal model parameters.

Moreover, data-driven methods have brought the significant "model-free" characteristic to power system operation. With a model-free approach, the distribution network operator does not have to collect circuit information and build individual models for each feeder which outperforms traditional model-based or hybrid approach with the shortcoming of reliance on circuit models [25]. Several data-driven methods are reported to resolve the model parameter mismatch or limited measurement issues. A model-free dynamic voltage control method for inverter-based energy resources is reported in [17] where pseudo gradients are proposed to adaptively approximate the optimal intermediate parameters. A reduced steady-state model synthesis method is presented in [26] to make advantage of dynamic measurements from phasor measurement units (PMUs) with a Kalman filter technique for data preparation. [27] proposes a data generation method for network reconfiguration then the desired result is solved by an improved convolution neural network trained by the generated data. [28] approximates power flow sensitivity model in second-order manner based on measurements to facilitate regulating voltage within reasonable range. [29] exploits the variability at metered buses and the stationarity of conventional loads to jointly solve the non-linear power flow equations then enhance the observability of ADNs. In [30], a joint estimation method to determine admittance parameters and topology in polyphase distribution network via lasso regression approach is presented. Nevertheless, aforementioned model identification methods are non-iterative thus are not suitable to be implemented for closed-loop control such as VVC.

In this paper, a data-driven VVC method is proposed, which is model-free. Firstly, a recursive online regression closed-from solution is proposed to approximate the real-time system input-output (or control-state) response combining the merits of decaying memory and limited memory of partial observations in ADNs. Then the recursive solution is embedded in the iterative data-driven optimization scheme to form a unified real-time VVC framework where the convergence of corresponded algorithm is proved. Abundant numerical tests exhibit the precise tracking performance of the proposed method approaching the optimal results solved based on the ideally accurate models. The adaptive learning property of the presented recursive regression methods is validated as well. The proposed strategy provides a near-global optimum in a real-time manner considering active power loss and voltage regulation.

The main contributions are summarized as follows.

1) Simply relied on partial observations in ADNs, a recursive regression close-form solution is proposed combing decaying memory and limited memory to update approximated system response policy in real-time.

2) The proposed data-driven optimization algorithm successfully pursues the global optimum with Q-linearly convergence.

3) Global optimization and local control are unified in the presented framework making the best advantage of rapid-switching inverter-based DERs.

The remainder of this paper is organized as follows. Section II introduces the time-varying property of power system and linear approximation of power flow as preliminaries for the following analysis. Then the online recursive regression method to dynamically estimate the linear power flow is demonstrated in Section III. Section IV presents the overall data-driven VVC strategy, along with theoretical analysis. Section IV presents numerical tests on unbalanced multi-phase distribution network case. Numerical studies are conducted in Section V to validate the effectiveness of the presented data-driven VVC method. Finally, Section VI concludes the paper and discusses about future extensions.

## II. Preliminaries

### A. Time-Varying Power System

A $N$-node three-phase distribution network described by directed graph graph $\Pi(\mathcal{N},\mathcal{L})$ with node set $\mathcal{N}$ and branch set $\mathcal{L}$:=$\{(i,j)\}\subset\mathcal{N}\times\mathcal{N}$ is considered where the slack bus is located at node $1$. Denote phase set as $\Gamma$=$\{\varphi\,|\,\varphi=A,\ B,\ C\}$. Distributed generators (DGs) are interfaced with the grid via smart inverters, denoted as set $\mathcal{G}$. The problem is described as a discrete-time model with a finite horizon of a day where time slots are indexed as $t=\{1,...,T\}$ with fixed time interval $\Delta t$.

Conventionally, the optimization problem of OPF variants is static which pursues a converged solution at each time slot. Nevertheless, for real-time implementation, system model is time-varying, described as

$$\begin{aligned} &\min_{\mathbf{x}_t,\mathbf{y}_t}\ F(\mathbf{x}_t,\mathbf{y}_t)\\ &\text{s.t.}\ \ \mathbf{y}_t=h(\mathbf{x}_t,\mathbf{r}_t,\mathbf{l}_t)\\ &\qquad g(\mathbf{x}_t,\mathbf{y}_t)\le 0 \end{aligned} \tag{1}$$

where $F$ refers to the target such as network loss, generation cost, etc. Power flow equation is described by the nonlinear mapping $h(\bullet)$ and operational constraints are concluded in $g(\bullet)$. $\mathbf{x}_t$ indicates control variables such as DG output, $\mathbf{y}_t$ is state variables like voltage magnitude, $\mathbf{r}_t$ is system model parameters like branch resistance or reactance, and $\mathbf{l}_t$ is the load parameters.

The optimum variation of the time-varying model is assumed to be restrained, demonstrated in Assumption 1.

**Assumption 1.** Denote $\mathbf{x}_t^*$ as the optimal solution of (1) at time *t*, then given a non-negative scalar $\theta_s$, the dynamic variation of the system is bounded as

$$\left\|\mathbf{x}_{t+1}^* - \mathbf{x}_t^*\right\|_2 \le \theta_s \tag{2}$$

Traditional static OPF solvers assume the accurate model parameters, which is not applicable for model (1) in ADNs. Moreover, loads are usually nonlinear in ADNs which face difficulties in recognizing detailed parameters as well.

*B. Power Flow Linear Approximation*

Although original power flow is nonlinear, there are multiple researches providing efficient linear approximation techniques of power flow to simplify the model, which develop rapid solutions for the time-varying problem and implement real-time OPF optimizer [20], [21]. However, loads in distribution networks are usually voltage dependent which impedes the online implementation. The nonlinearity of loads is also considered and linearly relaxed in [10] based on ZIP load model. Therefore, it is appropriate to rewrite $h(\bullet)$ in the following linear form as

$$\mathbf{y}_t \approx \mathbf{W}_t^{\mathrm{T}} \times \mathbf{x}_t + \mathbf{e}_t \tag{3}$$

where $\mathbf{W}_t$ and $\mathbf{e}_t$ are given as constants based on line parameters, load parameters and nominal operation points in previous literatures [32], [33]. $(\bullet)^{\mathrm{T}}$ is the transpose operator.

Nevertheless, as discussed before, $\mathbf{W}_t$ and $\mathbf{e}_t$ are not able to be acquired accurately thus a data-driven method is presented in section III as an alternative solution.

## III. Online Recursive Regression

In reality, the ADNs cannot be guaranteed to be fully observable since of limited real-time measurements. It is assumed that measurements are only available at DG integrated nodes and the point of common coupling (PCC) node, denoted as $\mathcal{M}$. Certainly, other critical nodes without DG integration can also be observable based on reality which provides more information for online control policy estimation. Assume there are *M* measurements and *G* DGs in total.

Therefore, the linear approximation power flow mapping (thereafter, it is named as the system response policy) is formulated as

$$\mathbf{y}_t \approx \widehat{\mathbf{W}}_t^{\mathrm{T}} \times \hat{\mathbf{x}}_t \tag{4}$$

where $\widehat{\mathbf{W}}_t \in \mathbf{R}^{(\mathrm{G}+1)\times \mathrm{M}}$ is the parameter matrix to be estimated, $\mathbf{y}_t \in \mathbf{R}^{\mathrm{M}}$ is the measurement column vector and the auxiliary column vector $\hat{\mathbf{x}}_t = \left[\mathbf{x}_t^{\mathrm{T}}, 1\right]^{\mathrm{T}} \in \mathbf{R}^{\mathrm{G}+1}$ is introduced for simplicity.

Then, it is reasonable to have the following assumption on the accuracy of the linear approximation.

**Assumption 2.** Linear regression errors compared with system response are bounded as

$$\left\|\mathbf{y}_t - \widehat{\mathbf{W}}_t^{\mathrm{T}} \times \hat{\mathbf{x}}_t\right\|_2 \le \theta_v \tag{5}$$

**Assumption 3.** Since the input data are limited in practice thus regression parameters are bounded as well.

$$\left\|\widehat{\mathbf{W}}_t\right\|_2 \le \theta_w \tag{6}$$

It is assumed that the system is changing slowly compared with the sample time interval as Assumption 1 indicates. In other words, within a given time window $L$, $\widehat{\mathbf{W}}_{t-L+1} \approx \cdots \approx \widehat{\mathbf{W}}_t \ (t \ge L)$ are replaced with $\widehat{\mathbf{W}}_{t-L+1,t}$ which implies only the limited memory of input data should be considered.

Moreover, the latest data contain the most accurate system response policy thus compared with outdated data, current data is more effective for linear approximation, namely the decaying memory.

For $t \ge L$, define $\mathbf{Y}_{t-L+1,t} = \left[\mathbf{y}_{t-L+1}, \cdots, \mathbf{y}_t\right] \in \mathbf{R}^{\mathrm{M}\times \mathrm{L}}$ and $\mathbf{X}_{t-L+1,t} = \left[\hat{\mathbf{x}}_{t-L+1}, \cdots, \hat{\mathbf{x}}_t\right] \in \mathbf{R}^{(\mathrm{G}+1)\times \mathrm{L}}$ then one gets the following regression model as

$$\min_{\widehat{\mathbf{W}}_{t-L+1,t}} (\mathbf{Y}_{t-L+1,t} - \widehat{\mathbf{W}}_{t-L+1,t}^{\mathrm{T}} \mathbf{X}_{t-L+1,t}) \mathcal{B}_L (\mathbf{Y}_{t-L+1,t} - \widehat{\mathbf{W}}_{t-L+1,t}^{\mathrm{T}} \mathbf{X}_{t-L+1,t})^{\mathrm{T}} \tag{7}$$

where diagonal matrix $\mathcal{B}_L \in \mathbf{R}^{\mathrm{L}\times \mathrm{L}}$ is formulated by decaying coefficient $0 \ll \beta < 1$ as

$$\mathcal{B}_L = \begin{bmatrix} \beta^{L-1} & & & & \\ & \ddots & & & \\ & & \beta^{L-m} & & \\ & & & \ddots & \\ & & & & 1 \end{bmatrix}, m = 1, \ldots, L \tag{8}$$

Denote $\widehat{\mathbf{X}}_{t-L+1,t} = \mathbf{X}_{t-L+1,t}\mathcal{B}_L^{\frac{1}{2}}$ and $\widehat{\mathbf{Y}}_{t-L+1,t} = \mathbf{Y}_{t-L+1,t}\mathcal{B}_L^{\frac{1}{2}}$ then the closed form solution of (7) is given as

$$\widehat{\mathbf{W}}_{t-L+1,t} = (\widehat{\mathbf{X}}_{t-L+1,t} \widehat{\mathbf{X}}_{t-L+1,t}^{\mathrm{T}})^{-1} \widehat{\mathbf{X}}_{t-L+1,t} \widehat{\mathbf{Y}}_{t-L+1,t}^{\mathrm{T}} \tag{9}$$

Nevertheless, the inverse operator is time-consuming for online implementation.

Recall the time-series property of input data and it is promising to develop a recursive solution. The intuition is to modify the parameter matrix in the last time slot by absorbing the latest measurements and discarding the most outdated measurements.

Define $\mathbf{\Phi}_{t-L+1,t} = (\widehat{\mathbf{X}}_{t-L+1,t} \widehat{\mathbf{X}}_{t-L+1,t}^{\mathrm{T}})^{-1}$. Then based on Sherman-Morrison formula, the following recursive formulation is given $(t \ge L+1)$.

$$\mathbf{\Phi}_{t-L,t} = \frac{1}{\beta}(\mathbf{I} - \frac{\mathbf{\Phi}_{t-L,t-1} \hat{\mathbf{x}}_t \hat{\mathbf{x}}_t^{\mathrm{T}}}{\beta + \hat{\mathbf{x}}_t^{\mathrm{T}} \mathbf{\Phi}_{t-L,t-1} \hat{\mathbf{x}}_t}) \mathbf{\Phi}_{t-L,t-1} \tag{10}$$

$$\mathbf{\Phi}_{t-L+1,t} = (\mathbf{I} + \frac{\beta^L \mathbf{\Phi}_{t-L,t} \hat{\mathbf{x}}_{t-L} \hat{\mathbf{x}}_{t-L}^{\mathrm{T}}}{1 - \beta^L \hat{\mathbf{x}}_{t-L}^{\mathrm{T}} \mathbf{\Phi}_{t-L,t} \hat{\mathbf{x}}_{t-L}}) \mathbf{\Phi}_{t-L,t} \tag{11}$$

where $\mathbf{I}$ indicates the identity matrix.

Then the recursive solution for $\widehat{\mathbf{W}}_{t-L+1,t}$ is given as

$$\widehat{\mathbf{W}}_{t-L+1,t}=\widehat{\mathbf{W}}_{t-L,t-1}+\mathbf{\Phi}_{t-L+1,t}\times\Delta \tag{12}$$

where augmented regression error matrix $\Delta$ is defined as $\Delta=\hat{\mathbf{x}}_t(\mathbf{y}_t^{\mathrm{T}}-\hat{\mathbf{x}}_t^{\mathrm{T}}\widehat{\mathbf{W}}_{t-L,t-1})-\beta^L\hat{\mathbf{x}}_{t-L}(\mathbf{y}_{t-L}^{\mathrm{T}}-\hat{\mathbf{x}}_{t-L}^{\mathrm{T}}\widehat{\mathbf{W}}_{t-L,t-1})$.

In this way, the update policy for $\widehat{\mathbf{W}}_{t-L+1,t}$ only requires simple calculation thus is applicable for online implementation. Detailed derivation of (10)-(12) is provided in Appendix A.

For the initial value $\mathbf{w}_{1,L}^i$, to avoid the collinearity of measurements, a positive scalar parameter $\lambda$ is introduced for the initial regression step:

$$\widehat{\mathbf{W}}_{1,L}=(\widehat{\mathbf{X}}_{1,L}\widehat{\mathbf{X}}_{1,L}^{\mathrm{T}}+\lambda\mathbf{I})^{-1}\widehat{\mathbf{X}}_{1,L}\widehat{\mathbf{Y}}_{1,L}^{\mathrm{T}} \tag{13}$$

## IV. Data-driven Online VVC

### A. Inverter-based DGs

We assume the inverter-based DGs are PVs operating at MPPT (maximum power point tracking) mode and this assumption is practical since most voltage violation issues are caused by distributed PVs in ADNs. Inverters are able to provide flexible reactive power support rapidly especially compared with legacy devices. Therefore, this advantage is leveraged as the critical physical condition for the online optimization. The operation constraints of inverters are defined as

$$\left|Q_{Gi,t}^{\varphi}\right|\le\sqrt{(S_{Gi}^{\varphi})^2-(\overline{P}_{Gi,t}^{\varphi})^2} \tag{14}$$

where $S_{Gi}^{\varphi}$ is the rating apparent power and $\overline{P}_{Gi}^{\varphi}$ is the ultra-short forecasted active output of PVs at corresponding phase. The feasible region of each inverter described by (14) is denoted as $\chi_i^t$. Furthermore, set $\chi^t:=\{\chi_i^t\mid i\in\mathcal{G}\}$ as the feasible domain of inverter output.

### B. Objective Function

To provide sufficient reactive power support, the voltage deviation compared with the predefined voltage profile $\hat{\mathbf{v}}_{\mathcal{M}}$ is the main concern for the rapid voltage regulation purpose. $\hat{\mathbf{v}}_{\mathcal{M}}$ can be predefined by the upper level stage in dispatch from distribution system operator (DSO).

Furthermore, the inverter loss should be considered since inverters operates frequently in the presented framework. One reports that the additional power losses of inverters caused by reactive power injection is proportional to $\left\|\mathbf{q}_{G,t}\right\|_2^2$ [34].

Meanwhile, system active power loss is decreased as the decrease of reactive power injection since fewer reactive injections potentially result in smaller line-current magnitude [8], [35].

Thus, the objective is to seek the optimal reactive power dispatch strategy to minimize both the voltage regulation cost and active power loss as

$$\min_{\mathbf{q}_G{}^t\in\chi^t} F_v=\frac{1}{2}(\alpha_1\left\|\mathbf{v}_{\mathcal{M}}^t\text{-}\hat{\mathbf{v}}_{\mathcal{M}}\right\|_2^2+\alpha_2\left\|\mathbf{q}_G{}^t\right\|_2^2) \tag{15}$$

where coefficient $\alpha_1$ and $\alpha_2$ indicate penalty of voltage regulation cost and active power loss, respectively. Moreover, these two coefficients can be extended as a series of coefficients so that each monitoring bus or DG have various weights. $\mathbf{v}_{\mathcal{M}}^t\in\mathbf{R}^{\mathrm{M}}$ imply measured voltage magnitudes at time $t$ at observable nodes as state variables and $\mathbf{q}_G{}^t\in\mathbf{R}^{\mathrm{G}}$ is the set of reactive power output as $\mathbf{q}_G{}^t:=\{Q_{Gi,t}^{\varphi}\mid i\in\mathcal{G},\varphi\in\Gamma\}$.

### C. Proposed VVC Strategy

By implementing the linear system response policy (4) as $\mathbf{v}_{\mathcal{M}}^t\approx\widehat{\mathbf{W}}_{t-L+1,t}^{\mathrm{T}}\times\begin{bmatrix}\mathbf{q}_G{}^t\\1\end{bmatrix}$ into (15), the overall optimization problem at each time slot $t$ is essentially approximated as a quadratic programming problem (16).

$$\begin{aligned}(VVC\text{ at time }t)\quad&\min_{\mathbf{q}_G{}^t} F_v=\frac{1}{2}(\alpha_1\left\|\mathbf{v}_{\mathcal{M}}^t\text{-}\hat{\mathbf{v}}_{\mathcal{M}}\right\|_2^2+\alpha_2\left\|\mathbf{q}_G{}^t\right\|_2^2)\\&\text{s.t.}\quad \mathbf{v}_{\mathcal{M}}^t\approx\widehat{\mathbf{W}}_{t-L+1,t}^{\mathrm{T}}\times\begin{bmatrix}\mathbf{q}_G{}^t\\1\end{bmatrix}\\&\mathbf{q}_G{}^t\in\chi^t\end{aligned} \tag{16}$$

Conventionally, an exact optimal solution for each time slot should be presented while external environment fluctuation is usually faster than the solution of converged optimum to the batched optimization [20].

Therefore, to provide the instant solution of VVC, inspired by real-time optimization [18]-[21], the online VVC strategy is presented where the online regression step and reactive power update step are integrated together in each iteration.

Specifically, the gradient-projection method, similar with the one in [36], is employed to provide the solution which unlike the conventional iterative method, the results of each iteration will be executed to the system instead of the final convergent solution.

Define the projection operator as $\mathcal{P}_{\chi}(\mathbf{x})=\underset{\mathbf{y}\in\chi}{\operatorname{argmin}}\left\|\mathbf{y}-\mathbf{x}\right\|_2$

Then the reactive power is updated with certain step size $d$ as

$$\mathbf{q}_G{}^{t+1}=\mathcal{P}_{\chi^t}(\mathbf{q}_G{}^t-d\times\nabla_{\mathbf{q}_G}F_v) \tag{17}$$

$$\begin{aligned}\nabla_{\mathbf{q}_G}F_v\,|_{\mathbf{q}_G{}^t}&=\alpha_1\frac{\partial\mathbf{v}_{\mathcal{M}}^t}{\partial\mathbf{q}_G}(\mathbf{v}_{\mathcal{M}}^t\text{-}\hat{\mathbf{v}}_{\mathcal{M}})+\alpha_2\mathbf{q}_G\,|_{\mathbf{q}_G{}^t}\\&=\alpha_1\widetilde{\mathbf{W}}_{t-L+1,t}\times(\mathbf{v}_{\mathcal{M}}^t\text{-}\hat{\mathbf{v}}_{\mathcal{M}})+\alpha_2\mathbf{q}_G{}^t\end{aligned} \tag{18}$$

where $\widetilde{\mathbf{W}}_{t-L+1,t}\in\mathbf{R}^{\mathrm{G}\times\mathrm{M}}$ represents $\widehat{\mathbf{W}}_{t-L+1,t}$ eliminating the las t row of $\widehat{\mathbf{W}}_{t-L+1,t}$.

Moreover, to alleviate the error influence brought the linear system response policy, real-time measurement $\tilde{\mathbf{v}}_{\mathcal{M}}^t$ is applied to directly present $\mathbf{v}_{\mathcal{M}}^t$ instead of calculating via (4). In other words, power flow solution is directly given by the physic law, which avoids the calculation process.

The critical part is that $\nabla_{\mathbf{q}_G}F_v$ is approximately replaced by $\Delta\mathbf{q}_G{}^t$ described in (20). (17) and (18) are rewritten as

$$\mathbf{q}_G{}^{t+1}=\mathcal{P}_{\chi^t}(\mathbf{q}_G{}^t-d\times\Delta\mathbf{q}_G{}^t) \tag{19}$$

$$\Delta\mathbf{q}_G{}^t=\alpha_1\widetilde{\mathbf{W}}_{t-L+1,t}(\tilde{\mathbf{v}}_{\mathcal{M}}^t\text{-}\hat{\mathbf{v}}_{\mathcal{M}})+\alpha_2\mathbf{q}_G{}^t \tag{20}$$

Since both $\widetilde{\mathbf{W}}_{t-L+1,t}$ and $\tilde{\mathbf{v}}^t_{\mathcal{M}}$ are given based on measurements, the presented approach is completely data-driven without any prior knowledge of the system model.

It should be also noted that after receiving $\Delta\mathbf{q}_G{}^t$ from control center, each DG updates its regulating target (19) locally.

### D. Data-driven VVC Framework

In this framework, the real-time measurements are continuously considered at each time slot as input data to correct the disturbance. Substantially, the proposed VVC method is a model-free data-driven approach. Detailed steps for the presented data-driven VVC method is listed in TABLE I.

TABLE I
ALGORITHM I DATA-DRIVEN ONLINE VVC STRATEGY

**Initialization:** Set $\mathbf{x}_t = \mathbf{q}_G{}^t$ and $\mathbf{y}_t = \tilde{\mathbf{v}}^t_{\mathcal{M}}$ . For, t=1,..L, control center collects measurement $\tilde{\mathbf{v}}^t_{\mathcal{M}}$ from critical nodes and $\mathbf{q}_G{}^t$ from DGs. Solve initial $\boldsymbol{w}^i_{1,L}$ within previous $L$ data window according to (13).

S1: **for** $t$=$L$+1,….. **do**

S2: Control center collects measurement $\mathbf{v}^t_{\mathcal{M}}$ from critical nodes and $\mathbf{q}_G{}^t$ from DGs.

S3: Control center updates $\widehat{\mathbf{W}}_{t-L+1,t}$ according to (12).

S4: Control center updates $\Delta\mathbf{q}_G{}^t$ according to (20).

S5: Control center sends $\Delta\mathbf{q}_G{}^t$ to corresponding DGs.

S6: DG regulate its reactive power $Q^{\varphi}_{Gi,t+1}$ locally according to (19).

S7: **end**

The overall VVC framework is illustrated in Fig. 1. The optimization in the ADN is iteratively solved with coordination between control center and DGs. Each iterative update will be applied by DG as commands to power system and the system measurements are input for the next iteration correction. Every step is essentially simple calculation thus the proposed VVC strategy is qualified for real-time implementation.

Usually, VVC is divided into two levels, primary and secondary control. In primary control, the local controller receives the command of reference point from secondary control to realize dynamic control and stabilize the real-time system operation like droop control. In secondary control, it is essentially a steady-state optimization stage which is this work. It should be noted that the proposed method is presenting reference point solution for Volt/Var optimization instead of local control. Therefore, the effective operation time resolution depends on the response capability of inverters.

More importantly, the presented scheme unifies the regression stage and optimization stage thus realizing the real-time optimization framework. In other words, global optimization and local control are unified in the proposed framework where each step the generated dispatch command is driving the system model pursuing global optimum at current time slot. The convergence of global optimal trajectory is proved below.

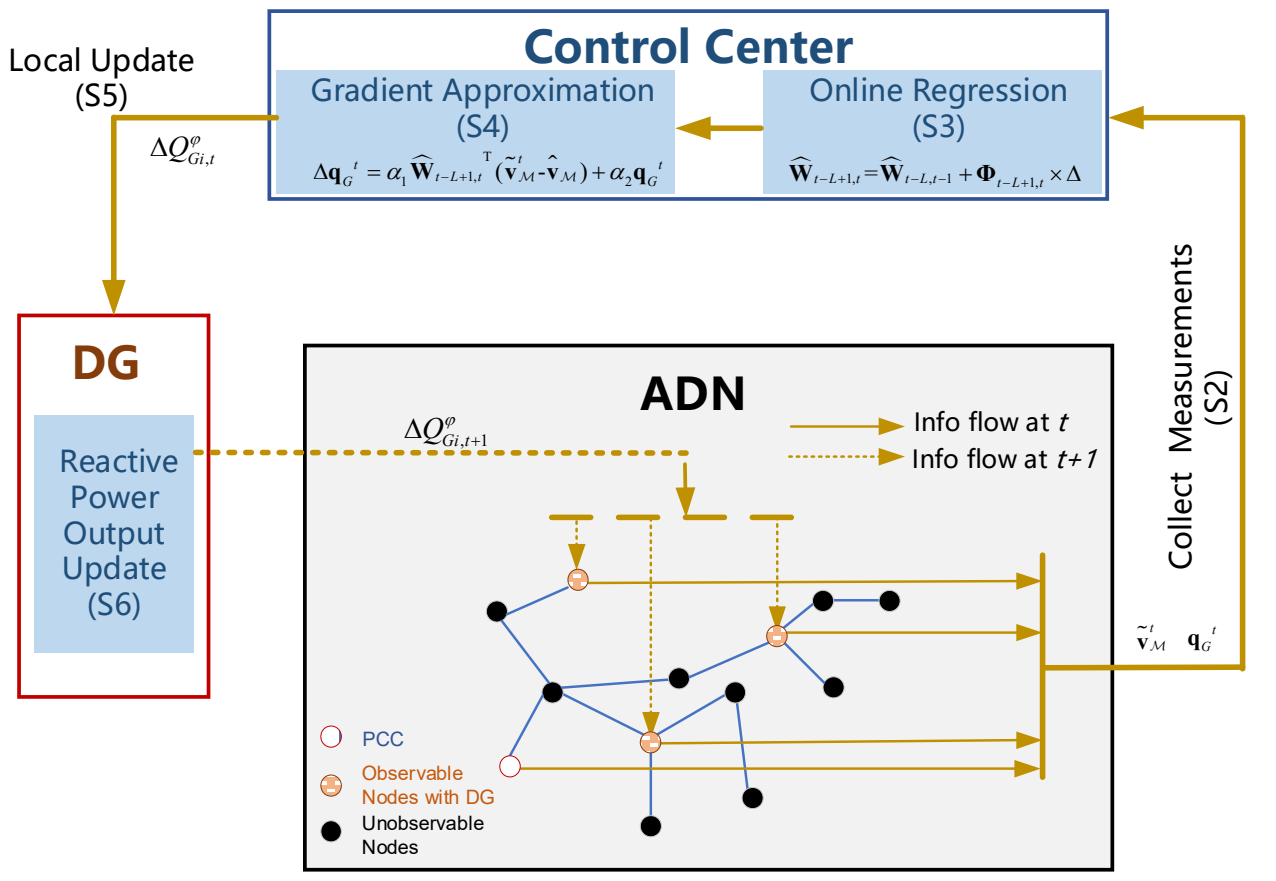

Fig. 1 Data-driven VVC framework.

### E. Convergence Analysis

Suppose Assumption 1, 2 and 3 hold then set $\mathbf{x}^{*t}_G = \mathbf{q}^{*t}_G$, $\mathbf{x}_t = \mathbf{q}_G{}^t$ and $\mathbf{y}_t = \mathbf{v}^t_{\mathcal{M}}$ . The convergence of the proposed strategy is demonstrated in Theorem 1.

**Theorem 1** Denote $\hat{\theta} = d\alpha_1\theta_v\theta_w + \theta_s$, if $0 < d < \dfrac{2}{\alpha_2}$, then $\mathbf{q}^t_G$ is Q-linearly convergent to the neighborhood of $\mathbf{q}^{*t}_G$ ($\hat{\theta} \neq 0$) or $\mathbf{q}^{*t}_G$ ($\hat{\theta} = 0$).

The proof is detailed in Appendix B.

## V. CASE STUDIES

Numerical simulations are conducted via MATLAB on a personal laptop with intel i7-8750 (2.2Ghz) and 16GB memory. A modified three-phase unbalanced IEEE 33-bus system are tested with system parameters obtained from [31]. It is assumed that only point of coupling (PCC) node and DG nodes are observable and the network outside the control area is unknown to the control center, illustrated in Fig. 2. Node 4 is set as PCC. Detailed PV configurations are listed in TABLE II. Nonlinear loads are assumed to be located at all phases and nodes. ZIP factors are obtained from [10] and randomly assigned to each load. The reference voltage at slack bus is set as 1 p.u.

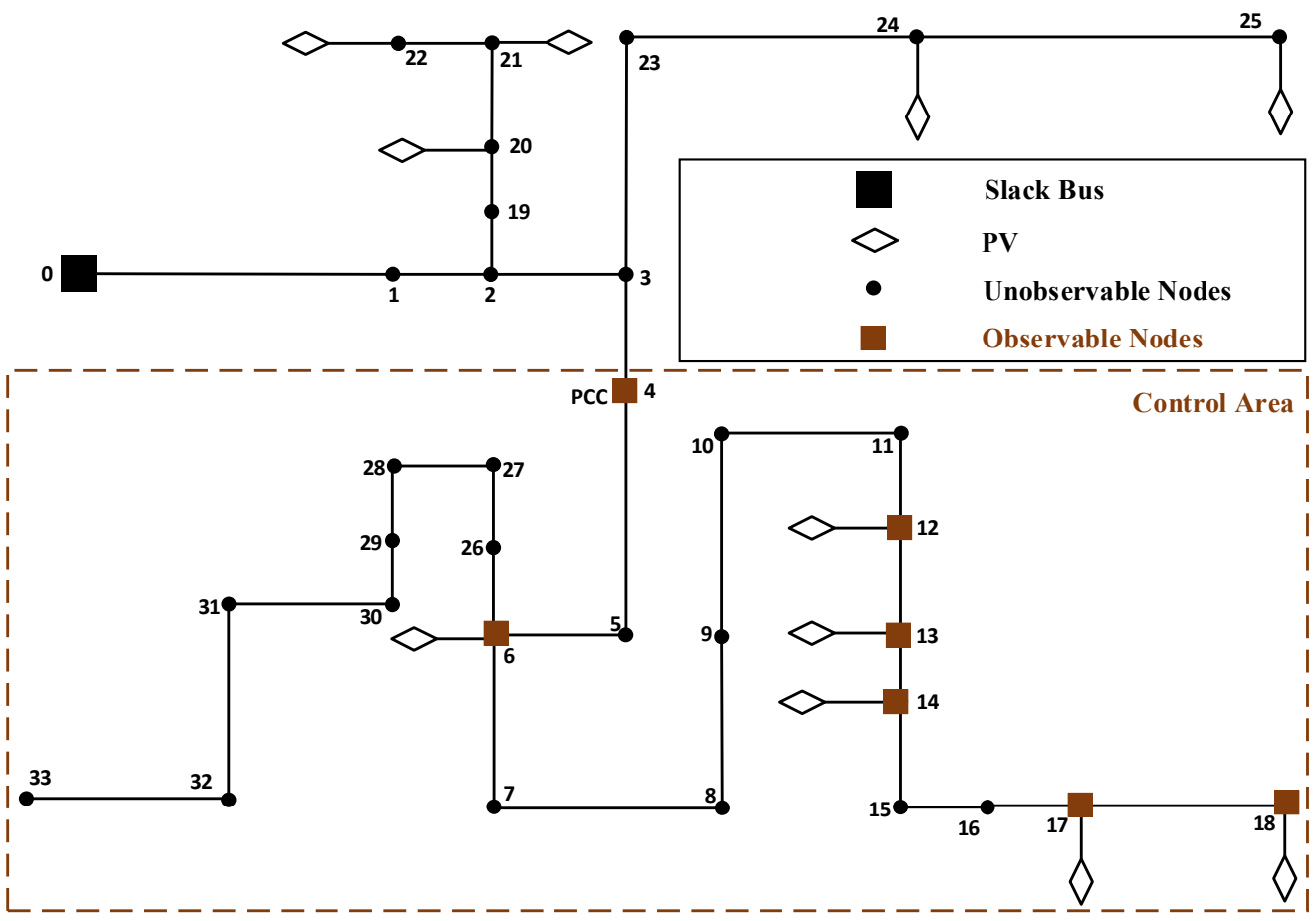

Fig. 2 Modified 33-bus system.

TABLE II
CONFIGURATION OF PV

| Integrated Node (Phase) | Capacity (kVA) |
|---|---|
| 12(ABC), 13(ABC), 21(AC), 25(BC) | 200 |
| 6(ABC), 14(BC), 20(AC), 24(AB), 26(AC) | 100 |
| 17(B), 18(B), 22(C) | 50 |

The normalized curves of total output of PVs and load demand are illustrated in Fig. 3 with the resolution of 1 second. The base MW is referred as the maximum value in the test system of PV and load respectively. There is a sudden change scenario starting at 10:00:15 shown in Fig. 3 (b) which is manually created for analysis later. Outside control area of PVs are assumed to operate at droop control mode.

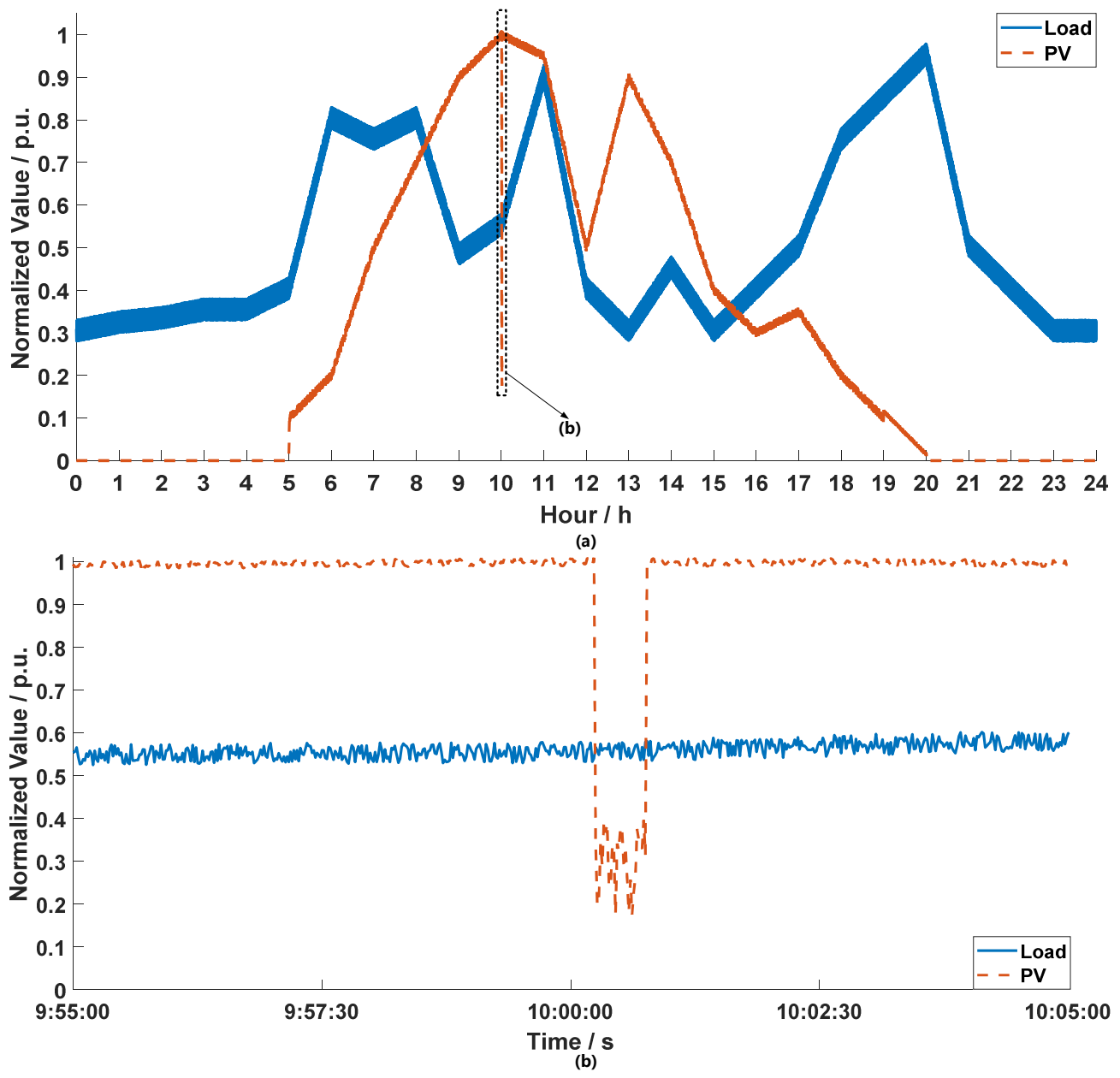

Fig. 3 PV and load normalized curve. (a) Overall curve. (b) Detailed curve at the neighborhood of the sudden change.

Gaussian noise with the mean value as 0 and variance as $0.001^2$ is added to the measurements. The implicit Zbus method [38] is integrated into simulations to present results of original nonlinear power flow considering nonlinear loads.

The control time interval is 1 second. Regression parameters are set as $L=10$ and $\beta=0.95$. Optimization parameters are set as $\alpha_1=10$, $\alpha_2=5$ and $d=0.1$ respectively.

## A. Linear Power Flow Approximation Accuracy

The proposed online regression method compares with the conventional model-based linear model [31]. Denote $\mathbf{v}^t_{\mathcal{M}}$ as the estimated voltage at time $t$ recovered by the linear approximation. Then the averaged error is defined by mean absolute error (MAE) as $V_{error}=\dfrac{\left\|\mathbf{v}^t_{\mathcal{M}}-\tilde{\mathbf{v}}^t_{\mathcal{M}}\right\|_1}{M}$.

As illustrated in Fig. 4, the typical time period when there suddenly is a cloud blocking sunlight is selected. It is shown that the proposed online regression method has less error at almost all time slots compared with the model-based linear method except for the 10:0:45. So, the proposed method has little bit worse performance during the sudden state change of the system explained by the "slowly changing" assumption mentioned in Section III. Nevertheless, the proposed method is obviously learning the power flow linear approximation by observing the improved accuracy after the sudden change. It is also because of the introduction of "decaying memory" mentioned in (8) and the effect that outdated data deteriorate the approximation performance is alleviated.

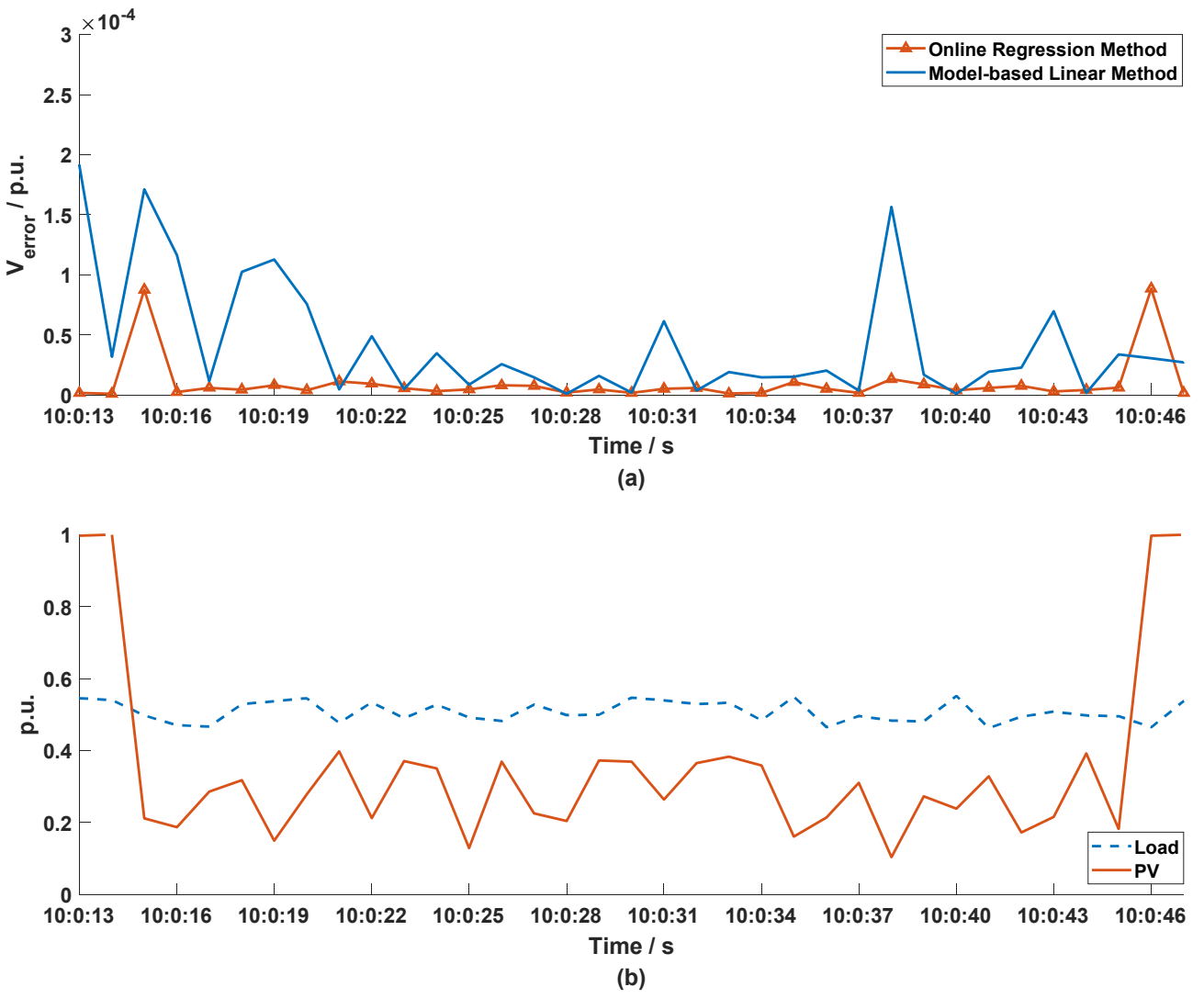

Fig. 4 Accuracy comparison between proposed online regression method and model-based method. (a) Approximation error. (b) PV and load profile.

More importantly, the parameters of the network and nonlinear loads are assumed to be accurate in the simulation for model-based benchmark while in reality those parameters are inaccurate or unknown. Therefore, the proposed online regression method significantly outperforms the regular model-based method both in accuracy and functionality.

## B. Comparison with Model-based Optimization Method

Then the comparison between the proposed data-driven method and model-based optimization method (converged solution to (1)) which is solved based on interior point solver is conducted at 10:0:17. The result in Fig. 5 indicates that the presented data-driven method successfully approaches the ideal optimal result, labeled as "model-based method (accurate)". On the contrast, the model-based optimization method delivers notable control result deviation under the inaccurate model scenario which is common in real practice. In this scenario, randomly disturbance of 20% model mismatch is added to each branch of the system. It indicates that the data-driven method is more practical than model-based method.

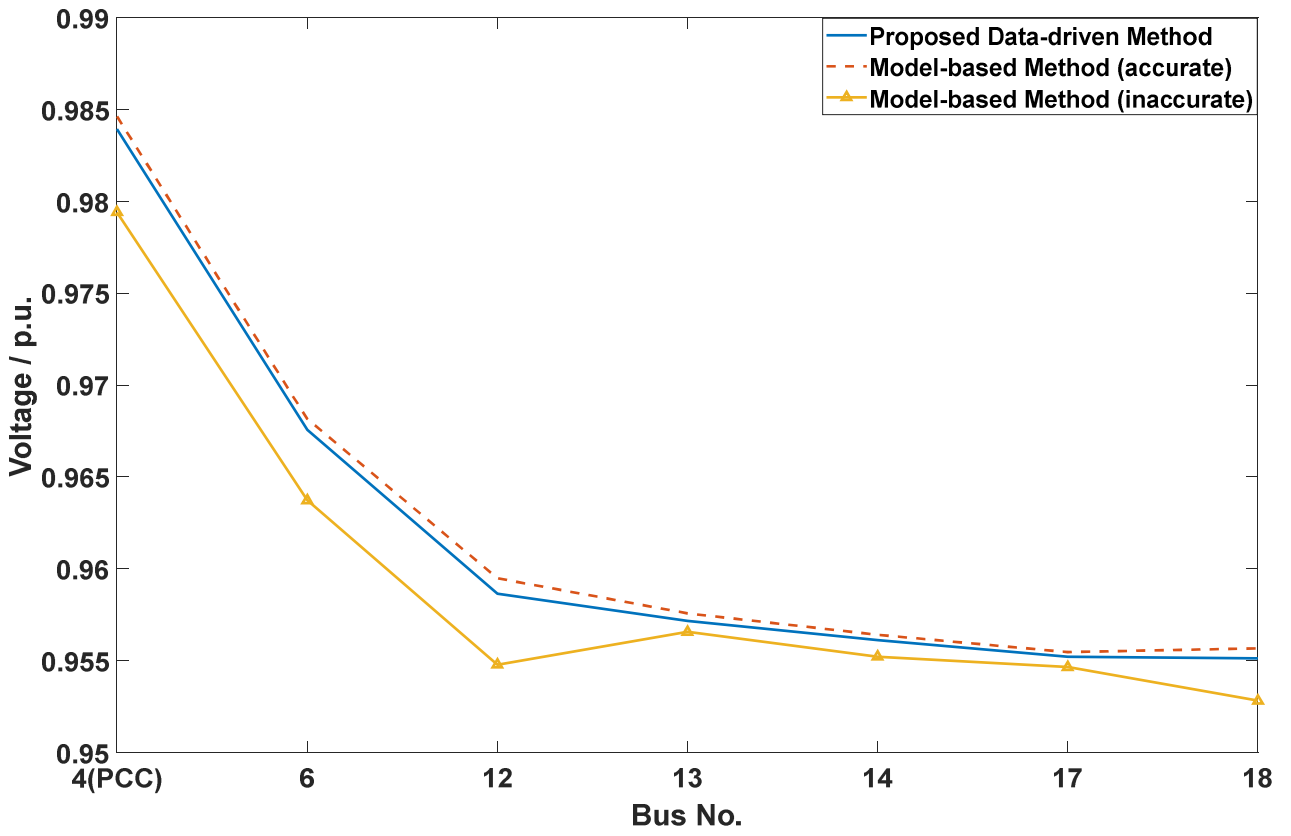

Fig. 5 Control result comparison between proposed online regression method and model-based method at 10:0:17.

## C. *Regulation Performance Comparison with Droop Control*

Droop control is a widely used strategy for PV local rapid control to regulate current local voltage magnitude $\widetilde{V}^t_{Gi}$ to stay close to the predefined value $\widehat{V}^t_{Gi}$ by setting the reactive power output update strategy as $Q^{\varphi}_{Gi,t+1} = \mathcal{P}_{\chi^t}\left[-\gamma_i \times \left(\widetilde{V}^t_{Gi} \text{-} \widehat{V}^t_{Gi}\right)\right]$ where $\gamma_i$ is a manually selected droop coefficient for each PV.

As shown in Fig. 6~Fig. 9, the proposed method exhibits better control performance with less network loss than those of the droop control. There are several voltage violations under droop control as well while the system operates within secure range under the proposed method. This is because the droop control is a pure local control strategy and ignores the global coordination. Similar result has been mentioned in [39] as well. Thus, the proposed data-driven method rapidly regulates voltage profile and simultaneously reduces the network loss.

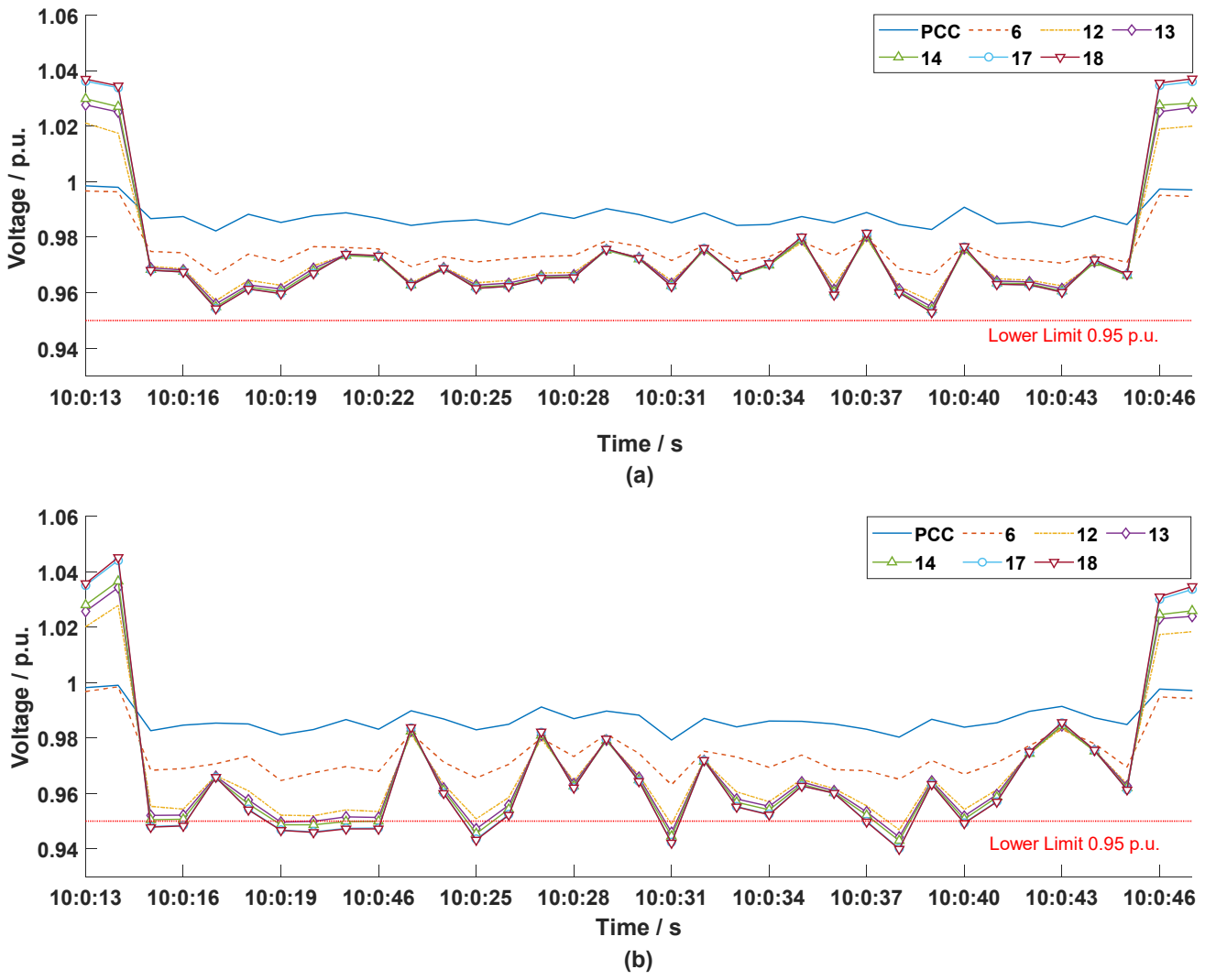

Fig. 6 Voltage regulation results comparison between proposed control and droop control (Phase A). (a) voltage profile of proposed control. (b) voltage profile of droop control.

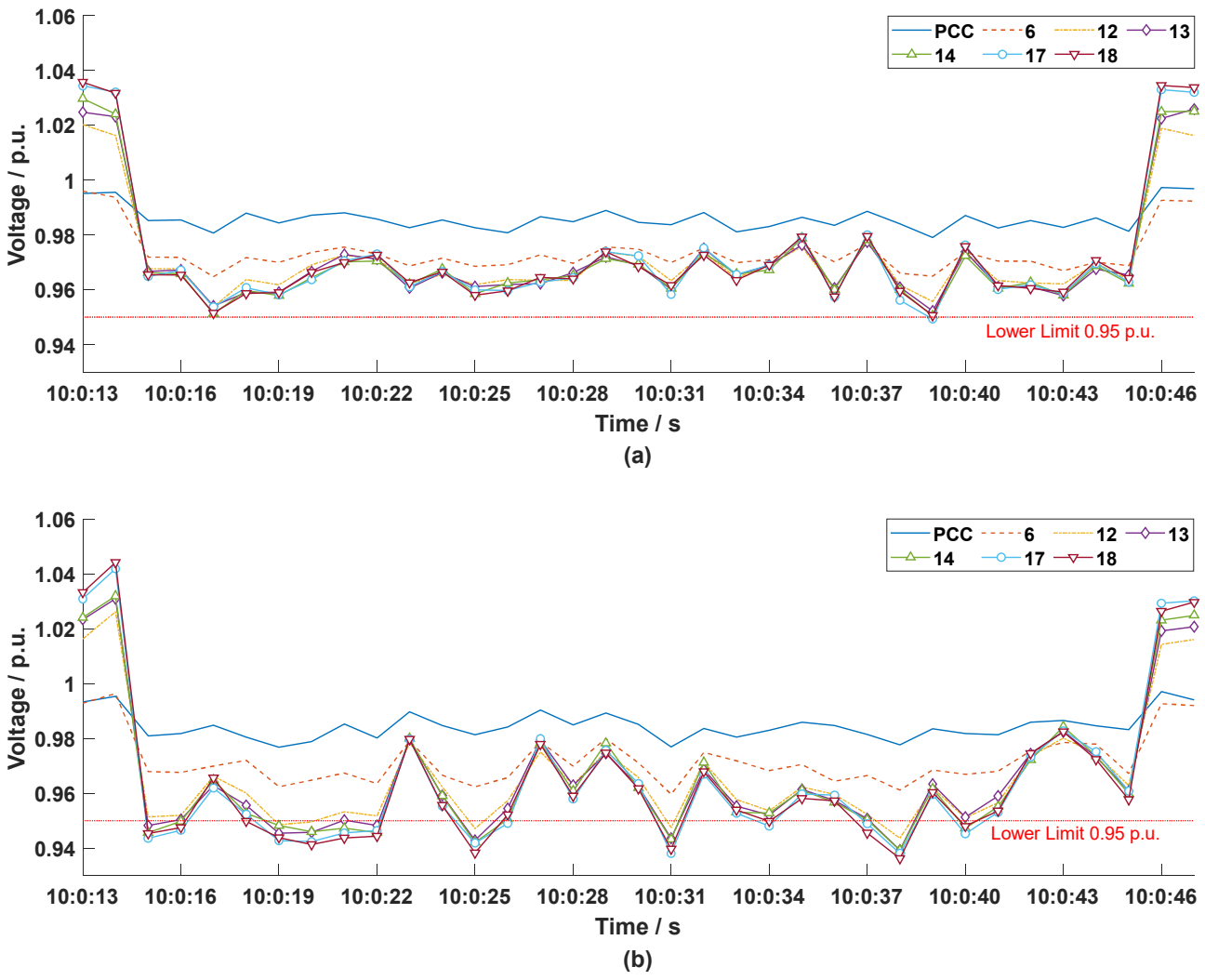

Fig. 7 Voltage regulation results comparison between proposed control and droop control (Phase B). (a) voltage profile of proposed control. (b) voltage profile of droop control.

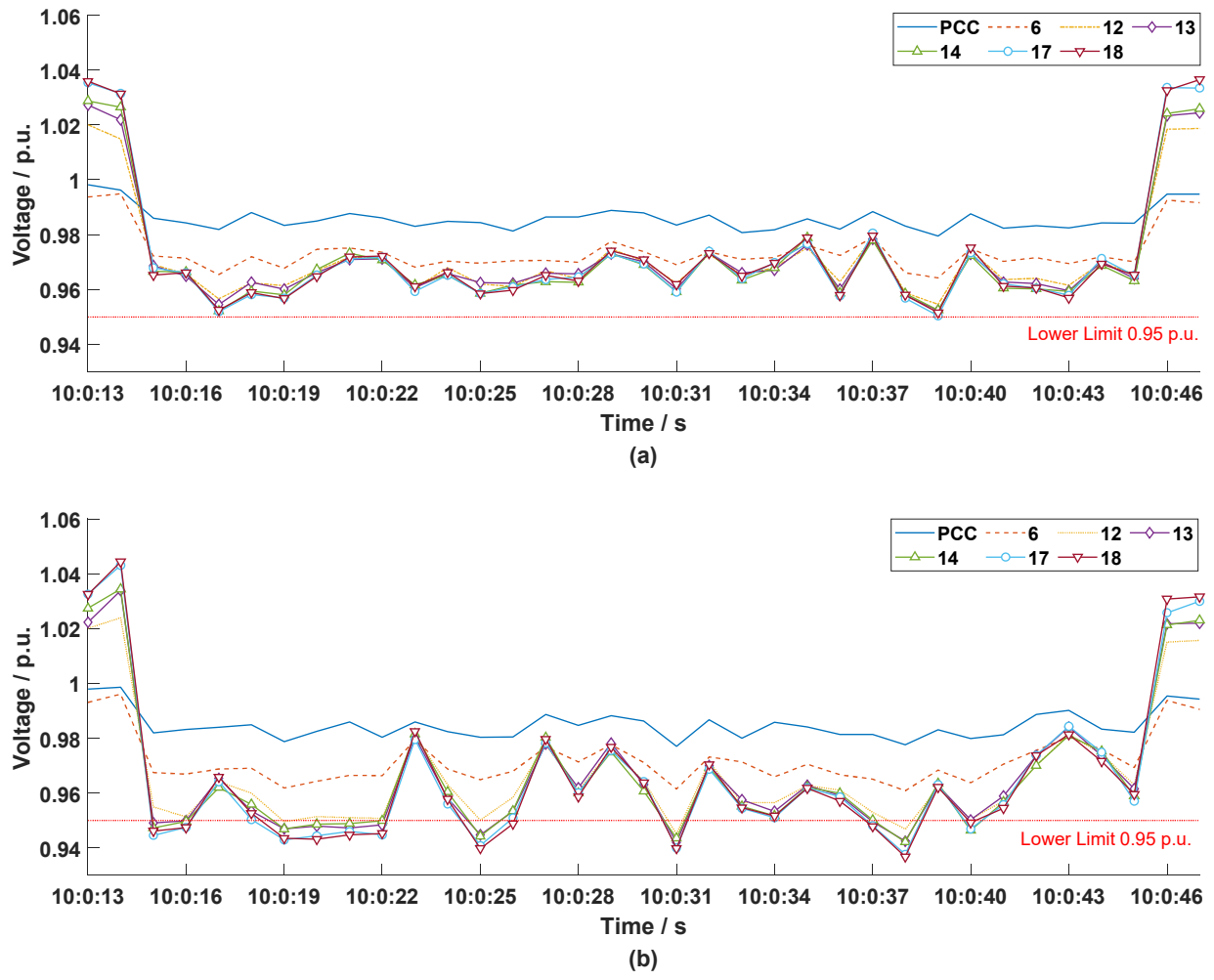

Fig. 8 Voltage regulation results comparison between proposed control and droop control (Phase C). (a) voltage profile of proposed control. (b) voltage profile of droop control.

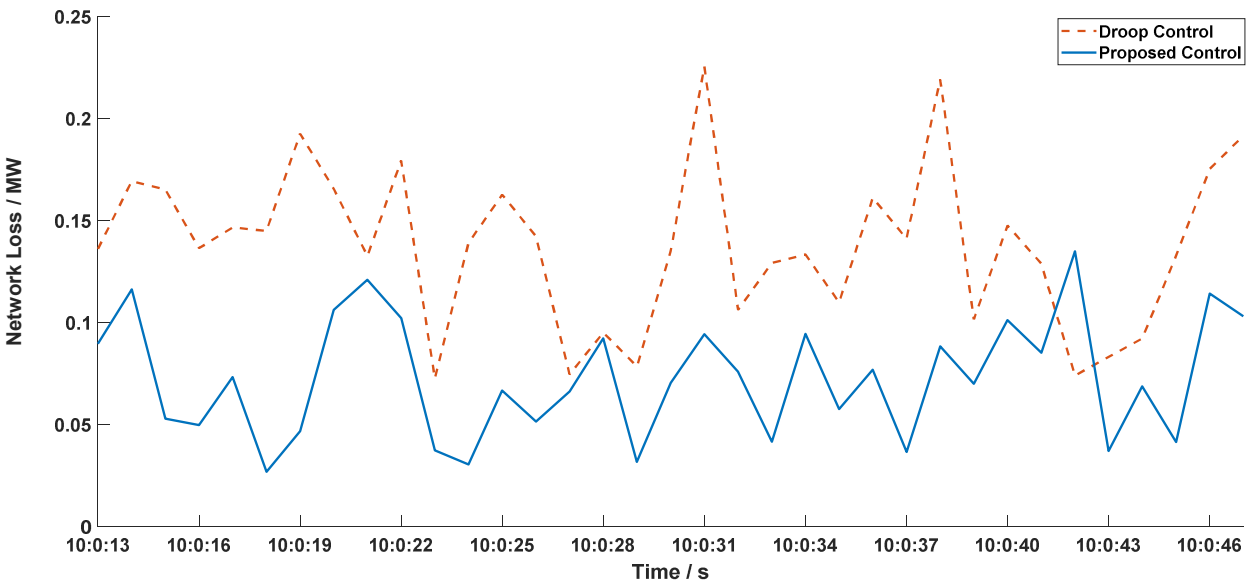

Fig. 9 Network active power loss comparison between proposed control and droop control (sum of three phases).

## D. *Effectiveness on Nonlinear Loads Case*

To validate the effectiveness of the proposed method against nonlinear loads, we assume loads are regular (specifically, “Z” and “I” coefficients of ZIP load model are zero) before 10:0:31 and nonlinear loads are integrated into all nodes of the system at 10:0:31. Take phase A for example, the result illustrated in Fig. 10~Fig. 12 implies that the presented approach is much more robust for voltage regulation when dealing with nonlinear loads compared with the conventional droop control method.

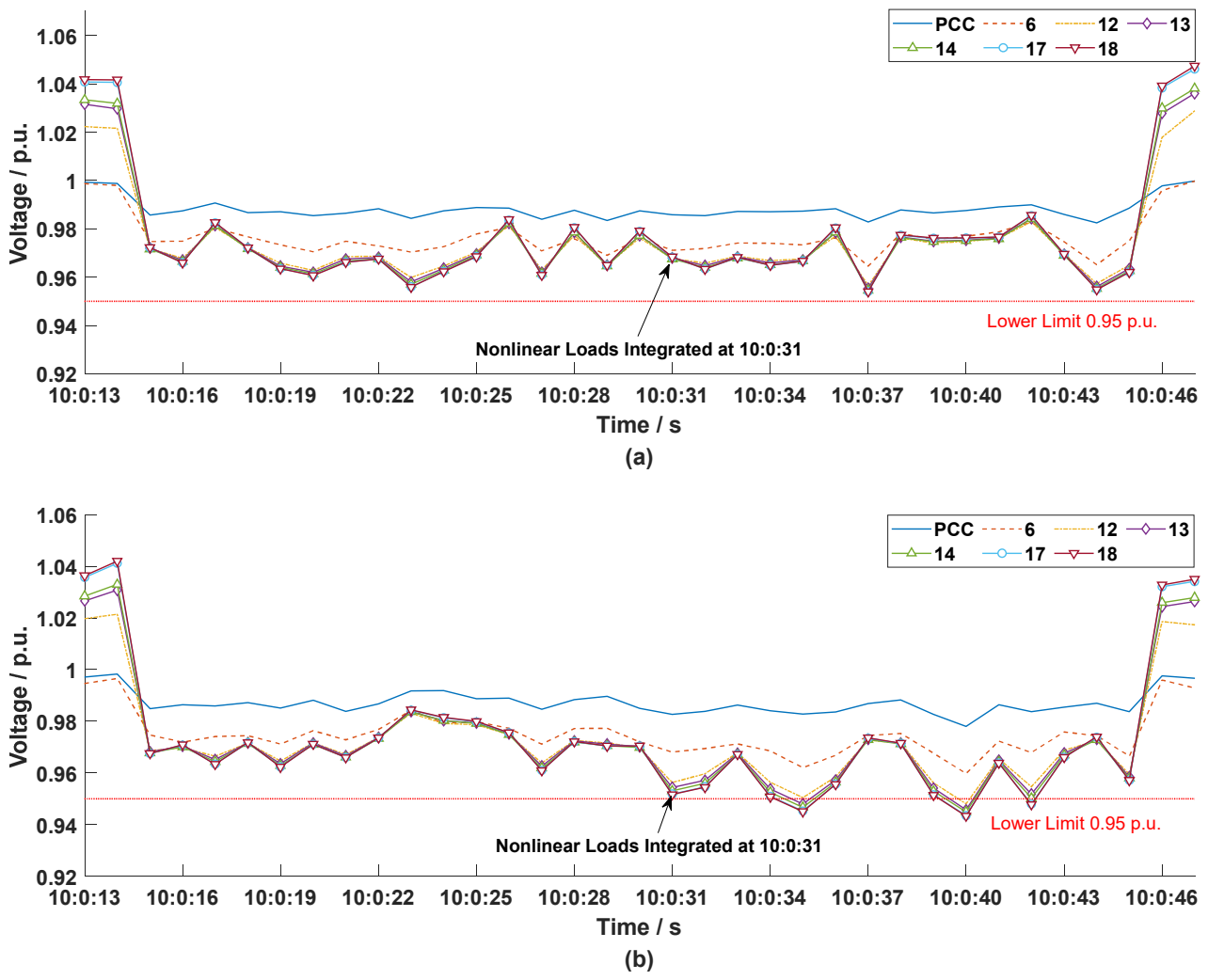

Fig. 10 Performance comparison between proposed control and droop control with nonlinear loads integrated at 10:0:31 (Phase A). (a) voltage profile of proposed control. (b) voltage profile of droop control.

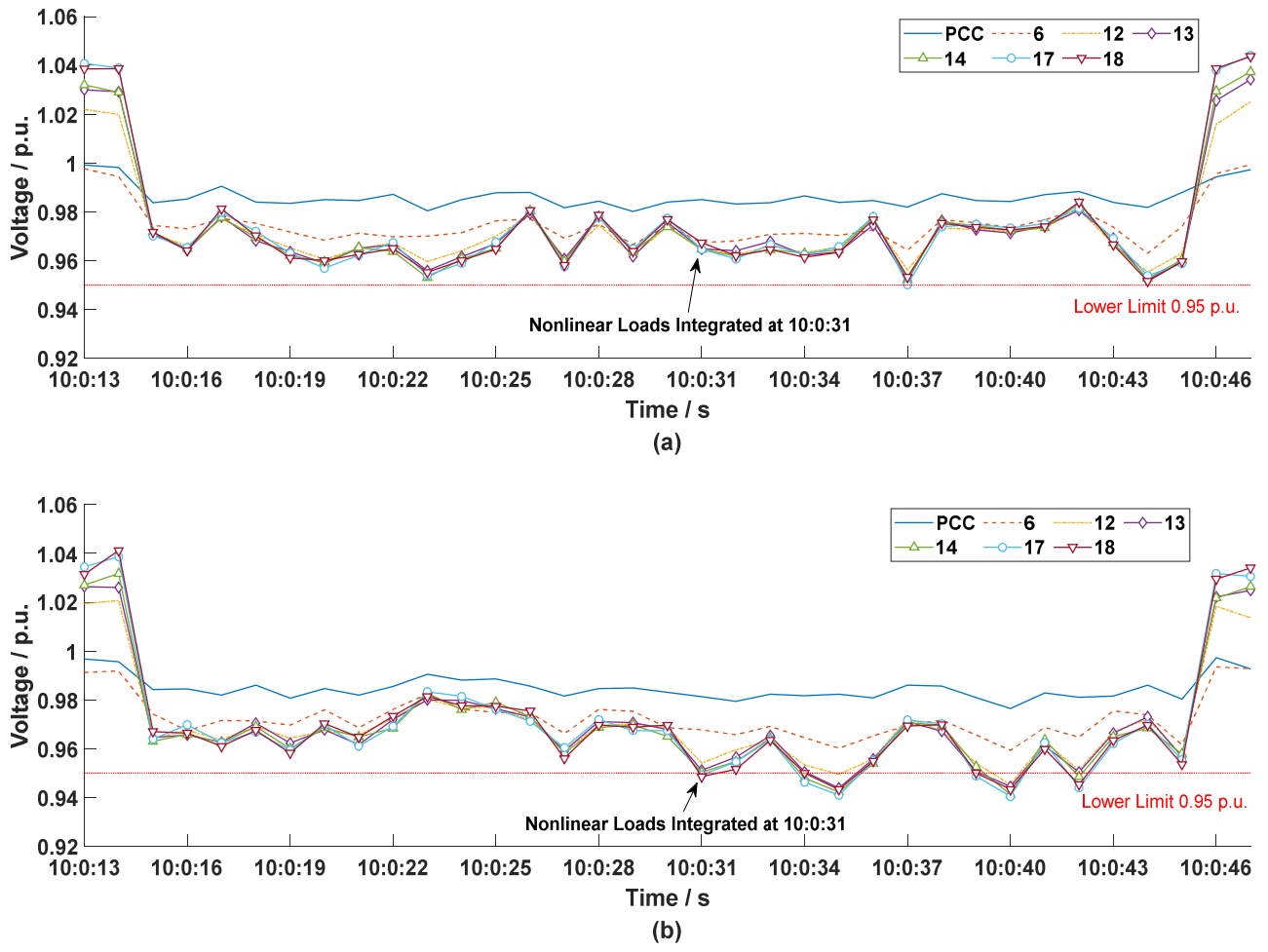

Fig. 11 Performance comparison between proposed control and droop control with nonlinear loads integrated at 10:0:31 (Phase B). (a) voltage profile of proposed control. (b) voltage profile of droop control.

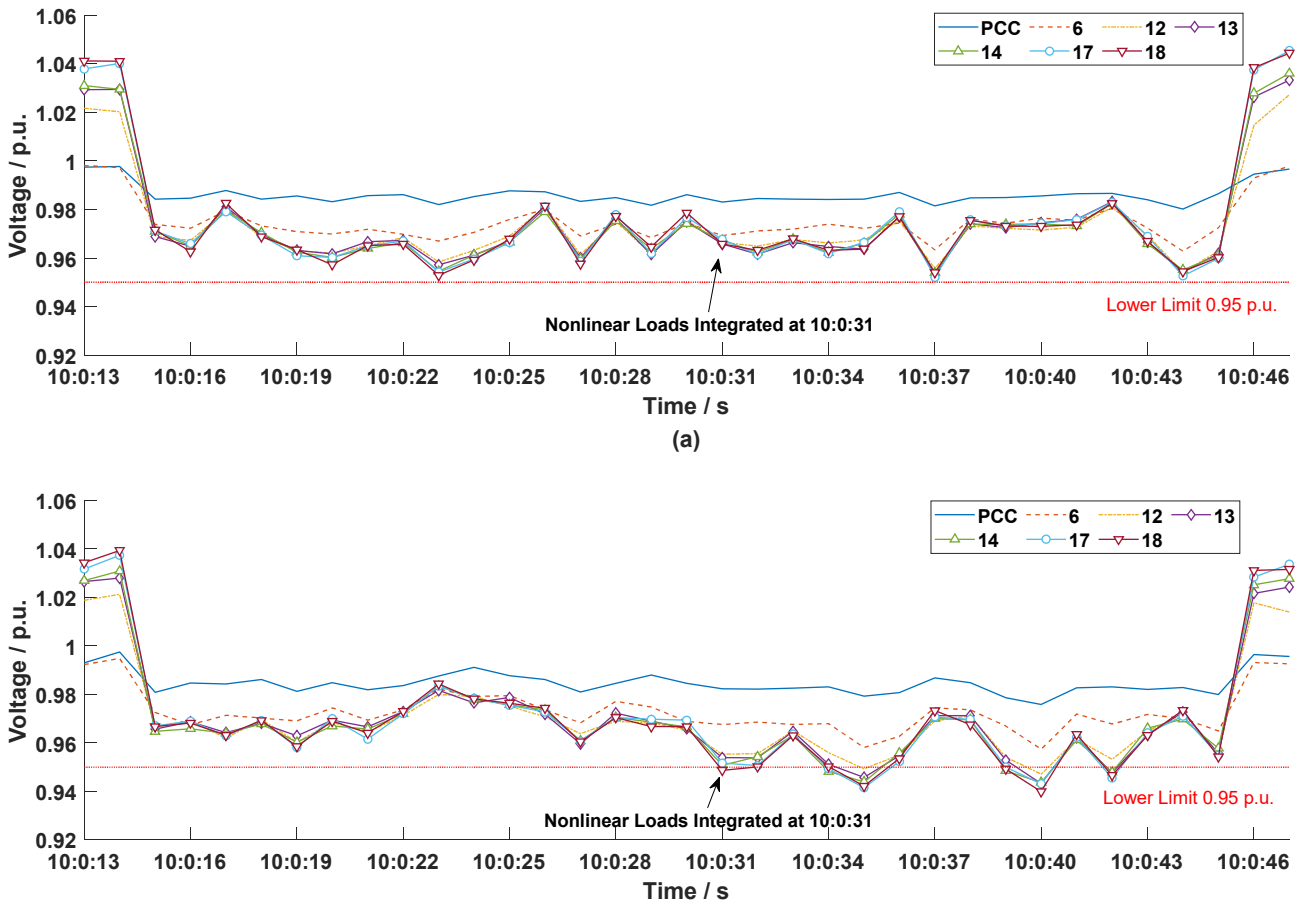

Fig. 12 Performance comparison between proposed control and droop control with nonlinear loads integrated at 10:0:31 (Phase C). (a) voltage profile of proposed control. (b) voltage profile of droop control.

### *E. Computation Burden*

The execution time of online regression update step and projection update step is 2.07ms and 0.89ms, respectively. The computation burden is very slight since the recursive regression algorithm is proposed, in which the inverse of matrix is only involved at the initial step. Therefore, it is practical to implement the proposed framework in real-time.

## VI. Conclusion

A data-driven model-free VVC method is presented in this paper which only relies partial measurements on critical nodes in ADNs. The proposed method approximates system response policy via a recursive regression algorithm then delivers correct voltage regulation strategies in real-time. The convergence of this VVC is proved theoretically and also verified by the numerical tests on an unbalanced three-phase ADNs. The numerical results show that the solutions at each time slot successfully approach the optimal points in iterative manner. And the presented method outperforms the conventional droop control under all circumstances. In this proposed closed-loop VVC framework, the computation burden is very slight, so it is suitable for real-time application. Future work may include extensions to fully distributed implementation. Moreover, stability analysis on the coordination between the proposed scheme and the inverter operation time shall be studied in the future as well.

## Appendix

### *A. Derivation of Recursive Solution of the Regression*

Since there is only fixed data window to be used, then the key is to “pull in” the latest data and “push out” the oldest data. Similar with $\mathbf{\Phi}_{t-L+1,t} = (\widehat{\mathbf{X}}_{t-L+1,t}\widehat{\mathbf{X}}_{t-L+1,t}^{\mathrm{T}})^{-1}$, there are definitions of $\mathbf{\Phi}_{t-L,t} = (\widehat{\mathbf{X}}_{t-L,t}\widehat{\mathbf{X}}_{t-L,t}^{\mathrm{T}})^{-1}$ and $\mathbf{\Phi}_{t-L,t-1} = (\widehat{\mathbf{X}}_{t-L,t-1}\widehat{\mathbf{X}}_{t-L,t-1}^{\mathrm{T}})^{-1}$.

According to Sherman-Morrison formula then (10) is derived as

$$\begin{aligned}\mathbf{\Phi}_{t-L,t} &= (\beta\widehat{\mathbf{X}}_{t-L,t-1}\widehat{\mathbf{X}}_{t-L,t-1}^{\mathrm{T}} + \hat{\mathbf{x}}_t\hat{\mathbf{x}}_t^{\mathrm{T}})^{-1} \\ &= \left[(\frac{\mathbf{\Phi}_{t-L,t-1}}{\beta})^{-1} + \hat{\mathbf{x}}_t\hat{\mathbf{x}}_t^{\mathrm{T}}\right]^{-1} \\ &= \frac{1}{\beta}(\mathbf{I} - \frac{\mathbf{\Phi}_{t-L,t-1}\hat{\mathbf{x}}_t\hat{\mathbf{x}}_t^{\mathrm{T}}}{\beta + \mathbf{x}_t^{\mathrm{T}}\mathbf{\Phi}_{t-L,t-1}\mathbf{x}_t})\mathbf{\Phi}_{t-L,t-1}\end{aligned} \tag{21}$$

Similar, (11) is also given by

$$\begin{aligned}\mathbf{\Phi}_{t-L+1,t} &= (\widehat{\mathbf{X}}_{t-L,t}\widehat{\mathbf{X}}_{t-L,t}^{\mathrm{T}} - \beta^{L}\hat{\mathbf{x}}_{t-L}\hat{\mathbf{x}}_{t-L}^{\mathrm{T}})^{-1} \\ &= (\mathbf{I} + \frac{\beta^{L}\mathbf{\Phi}_{t-L,t}\hat{\mathbf{x}}_{t-L}\hat{\mathbf{x}}_{t-L}^{\mathrm{T}}}{1-\beta^{L}\mathbf{x}_{t-L}^{\mathrm{T}}\mathbf{\Phi}_{t-L,t}\mathbf{x}_{t-L}})\mathbf{\Phi}_{t-L,t}\end{aligned} \tag{22}$$

Moreover, recall $\mathbf{\Phi}_{t-L,t-1}^{-1}\widehat{\mathbf{W}}_{t-L,t-1} = \widehat{\mathbf{X}}_{t-L,t-1}\widehat{\mathbf{Y}}_{t-L,t-1}^{\mathrm{T}}$ then $\widehat{\mathbf{W}}_{t-L,t}$ is recursively solved as follows.

$$\begin{aligned}\widehat{\mathbf{W}}_{t-L,t} &= \mathbf{\Phi}_{t-L,t}\widehat{\mathbf{X}}_{t-L,t}\widehat{\mathbf{Y}}_{t-L,t}^{\mathrm{T}}\\ &=\mathbf{\Phi}_{t-L,t}(\beta\mathbf{\Phi}_{t-L,t-1}^{-1}\widehat{\mathbf{W}}_{t-L,t-1}+\hat{\mathbf{x}}_t\mathbf{y}_t^{\mathrm{T}})\\ &=\mathbf{\Phi}_{t-L,t}\beta\frac{1}{\beta}(\mathbf{\Phi}_{t-L,t}^{-1}-\hat{\mathbf{x}}_t\hat{\mathbf{x}}_t^{\mathrm{T}})\widehat{\mathbf{W}}_{t-L,t-1}+\mathbf{\Phi}_{t-L,t}\mathbf{x}_t\mathbf{y}_t^{\mathrm{T}}\\ &=\widehat{\mathbf{W}}_{t-L,t-1}+\mathbf{\Phi}_{t-L,t}\mathbf{x}_t(\mathbf{y}_t^{\mathrm{T}}-\mathbf{x}_t^{\mathrm{T}}\widehat{\mathbf{W}}_{t-L,t-1})\end{aligned} \tag{23}$$

Similarly, based on $\widehat{\mathbf{W}}_{t-L,t}$, $\widehat{\mathbf{W}}_{t-L,t+1}$ is given as

$$\begin{aligned}\widehat{\mathbf{W}}_{t-L+1,t} &= \mathbf{\Phi}_{t-L+1,t}\widehat{\mathbf{X}}_{t-L+1,t}\widehat{\mathbf{Y}}_{t-L+1,t}^{\mathrm{T}}\\ &= \mathbf{\Phi}_{t-L+1,t}(\widehat{\mathbf{X}}_{t-L,t}\widehat{\mathbf{Y}}_{t-L,t}^{\mathrm{T}}-\beta^L\hat{\mathbf{x}}_{t-L}\mathbf{y}_{t-L}^{\mathrm{T}})\\ &=\mathbf{\Phi}_{t-L+1,t}(\mathbf{\Phi}_{t-L,t}^{-1}\widehat{\mathbf{W}}_{t-L,t}-\beta^L\hat{\mathbf{x}}_{t-L}\mathbf{y}_{t-L}^{\mathrm{T}})\\ &= \mathbf{\Phi}_{t-L+1,t}(\mathbf{\Phi}_{t-L+1,t}^{-1}+\beta^L\hat{\mathbf{x}}_{t-L}\hat{\mathbf{x}}_{t-L}^{\mathrm{T}})\widehat{\mathbf{W}}_{t-L,t}\\ &\quad-\mathbf{\Phi}_{t-L+1,t}\beta^L\mathbf{x}_{t-L}\mathbf{y}_{t-L}^{\mathrm{T}}\\ &=\widehat{\mathbf{W}}_{t-L,t}-\beta^L\mathbf{\Phi}_{t-L+1,t}\mathbf{x}_{t-L}(\mathbf{y}_{t-L}^{\mathrm{T}}-\mathbf{x}_{t-L}^{\mathrm{T}}\widehat{\mathbf{W}}_{t-L,t})\end{aligned} \tag{24}$$

Then combing (23) and (24), (12) is derived.

## B. *Proof of Theorem 1*

Due to the non-expansive property of projection mapping, in the following analysis, projection operator is omitted for simplicity. The convergence analysis is inspired by the work [20].

Denote $\theta^t = \alpha_1\widehat{\mathbf{W}}_{t-L+1,t}^{\mathrm{T}}\times(\mathbf{v}_{\mathcal{M}}^t-\tilde{\mathbf{v}}_{\mathcal{M}}^t)$ as the gradient error.

Then based on Assumption 2 and 3, one gets

$$\left\|\theta^t\right\|_2 \le \alpha_1\theta_v\theta_w \tag{25}$$

Furthermore, recall Assumption 1 and the following inequality holds.

$$\begin{aligned}\left\|\mathbf{q}_G^{t+1}-\mathbf{q}_G^{*t+1}\right\|_2 &= \left\|\mathbf{q}_G^{t+1}-\mathbf{q}_G^{*t}+\mathbf{q}_G^{*t+1}-\mathbf{q}_G^{*t}\right\|_2\\ &\le \left\|\mathbf{q}_G^{t+1}-\mathbf{q}_G^{*t}\right\|_2+\theta_s\\ &=\left\|\mathbf{q}_G^{t}-d\times\Delta\mathbf{q}_G^{t}-\mathbf{q}_G^{*t}+d\times\nabla_{\mathbf{q}_G}F_v\,|_{\mathbf{q}_G^{*t}}\right\|_2+\theta_s\\ &\le\left|1-\alpha_2 d\right|\left\|\mathbf{q}_G^{t}-\mathbf{q}_G^{*t}\right\|_2+d\left\|\theta^t\right\|_2+\theta_s\\ &\le\left|1-\alpha_2 d\right|\left\|\mathbf{q}_G^{t}-\mathbf{q}_G^{*t}\right\|_2+d\alpha_1\theta_v\theta_w+\theta_s\end{aligned} \tag{26}$$

Then if $|1-\alpha_2 d|<1$, (26) is a contraction and the proof is complete.

## REFERENCES

[1] C. Vartanian, R. Bauer, *et al.*, "Ensuring System Reliability: Distributed Energy Resources and Bulk Power System Considerations," *IEEE Power Energy Mag.*, vol. 16, no. 6, pp. 52-63, Nov.-Dec. 2018.

[2] Almasalma, Hamada, Sander Claeys, and Geert Deconinck. "Peer-to-peer-based integrated grid voltage support function for smart photovoltaic inverters." *Applied Energy*, vol.239, pp.1037-1048, 2019.

[3] Z. Li, H. Sun, Q. Guo, J. Wang and G. Liu, "Generalized Master–Slave-Splitting Method and Application to Transmission–Distribution Coordinated Energy Management," *IEEE Trans. Power Syst.*, vol. 34, no. 6, pp. 5169-5183, Nov. 2019

[4] J. Lin *et al.*, "Situation awareness of active distribution network: Roadmap, technologies, and bottlenecks," *CSEE J. Power Energy Syst.*, vol. 2, no. 3, pp. 35–42, Sep. 2016.

[5] W. Zheng, W. Wu, B. Zhang and Y. Wang, "Robust reactive power optimisation and voltage control method for active distribution networks via dual time-scale coordination, " *IET Gener., Transm. Distrib.*, vol. 11, no. 6, pp. 1461–1471, May 2017.

[6] H. J. Liu, W. Shi and H. Zhu, "Hybrid Voltage Control in Distribution Networks Under Limited Communication Rates," *IEEE Trans. Smart Grid*, vol. 10, no. 3, pp. 2416-2427, May 2019.

[7] IEEE PES Industry Technical Support Task Force, "Impact of IEEE 1547 Standard on Smart Inverters," IEEE, May, 2018.

[8] H. J. Liu, W. Shi and H. Zhu, "Distributed Voltage Control in Distribution Networks: Online and Robust Implementations," *IEEE Trans. Smart Grid*, vol. 9, no. 6, pp. 6106-6117, Nov. 2018.

[9] E. Dall'Anese, S. V. Dhople and G. B. Giannakis, "Photovoltaic Inverter Controllers Seeking AC Optimal Power Flow Solutions," *IEEE Trans. Power Syst.*, vol. 31, no. 4, pp. 2809-2823, July 2016

[10] R. R. Jha, A. Dubey, C. Liu and K. P. Schneider, "Bi-Level Volt-VAR Optimization to Coordinate Smart Inverters With Voltage Control Devices," *IEEE Trans. Power Syst.*, vol. 34, no. 3, pp. 1801-1813, 2019.

[11] D. K. Molzahn et al., "A Survey of Distributed Optimization and Control Algorithms for Electric Power Systems," *IEEE Trans. Smart Grid*, vol. 8, no. 6, pp. 2941-2962, Nov. 2017.

[12] J. W. Simpson-Porco, F. Dörfler, and F. Bullo, "Voltage stabilization in microgrids via quadratic droop control," IEEE Trans. Autom. Control, vol. 62, no. 3, pp. 1239–1253, Mar. 2017.

[13] H. Zhu and H. J. Liu, "Fast local voltage control under limited reactive power: Optimality and stability analysis," *IEEE Trans. Power Syst.*, vol. 31, no. 5, pp. 3794–3803, Sep. 2016.

[14] M. Todescato, J. W. Simpson-Porco, F. Dörfler, R. Carli and F. Bullo, "Online Distributed Voltage Stress Minimization by Optimal Feedback Reactive Power Control," *IEEE Trans. Control Netw. Syst.*, vol. 5, no. 3, pp. 1467-1478, Sept. 2018.

[15] S. Bolognani, R. Carli, G. Cavraro, and S. Zampieri, "Distributed reactive power feedback control for voltage regulation and loss minimization," *IEEE Trans. Autom. Control*, vol. 60, no. 4, pp. 966–981, Apr. 2015.

[16] A J Wood, B F Wollenberg, G B Sheblé. *Power generation, operation, and control*. USA: John Wiley & Sons, 2013.

[17] H. Liu, W. Wu and A. Bose, "Model-free Voltage Control for Inverter-based Energy Resources: Algorithm, Simulation and Field Test Verification," *IEEE Trans. Energy Convers.*, in press.

[18] Z. Wang, W. Wu and B. Zhang, "A Distributed Control Method With Minimum Generation Cost for DC Microgrids," *IEEE Transactions on Energy Conversion*, vol. 31, no. 4, pp. 1462-1470, Dec. 2016.

[19] Z. Wang, W. Wu and B. Zhang, "A Fully Distributed Power Dispatch Method for Fast Frequency Recovery and Minimal Generation Cost in Autonomous Microgrids," *IEEE Trans. Smart Grid*, vol. 7, no. 1, pp. 19-31, Jan. 2016.

[20] E. Dall'Anese and A. Simonetto, "Optimal Power Flow Pursuit," *IEEE Trans. Smart Grid*, vol. 9, no. 2, pp. 942-952, March 2018.

[21] Y. Tang, K. Dvijotham, and S. Low, "Real-time optimal power flow", *IEEE Trans. Smart Grid*, vol. 8, no. 6, pp. 2963–2973, Nov. 2017.

[22] G. Qu and N. Li, "Optimal distributed feedback voltage control under limited reactive power, " *IEEE Trans. Power Syst.*, vol. 35, no. 1, pp.315–331, Jan. 2020.

[23] A. Bernstein and E. Dall'Anese, "Real-Time Feedback-Based Optimization of Distribution Grids: A Unified Approach," *IEEE Trans. Control of Netw. Syst.*, vol. 6, no. 3, pp. 1197-1209, Sept. 2019.

[24] L. Ortmann, A. Hauswirth, I. Caduff, F. Dörfler, S. Bolognani, "Experimental validation of feedback optimization in power distribution grids," *Electric Power Systems Research*, Vol. 189, 2020.

[25] P. Bagheri and W. Xu, "Model-Free Volt-Var Control Based on Measurement Data Analytics," *IEEE Trans. Power Syst.*, vol. 34, no. 2, pp. 1471-1482, March 2019.

[26] F. Mahmood, H. Hooshyar, J. Lavenius, A. Bidadfar, P. Lund and L. Vanfretti, "Real-Time Reduced Steady-State Model Synthesis of Active Distribution Networks Using PMU Measurements," *IEEE Trans. Power Delivery*, vol. 32, no. 1, pp. 546-555, Feb. 2017.

[27] Z. Yin, X. Ji, Y. Zhang, Q. Liu and X. Bai, "Data-driven approach for real-time distribution network reconfiguration," *IET Gener., Transm. & Distri.*, vol. 14, no. 13, pp. 2450-2463, 3 7 2020.

[28] H. Su, P. Li, *et al.*, "Augmented sensitivity estimation based voltage control strategy of active distribution networks with PMU measurement," *IEEE Access*, vol. 7, pp. 44 987–44 997, Mar. 2019.

[29] S. Bhela, V. Kekatos and S. Veeramachaneni, "Enhancing Observability in Distribution Grids Using Smart Meter Data," *IEEE Trans. Smart Grid*, vol. 9, no. 6, pp. 5953-5961, Nov. 2018.

[30] O. Ardakanian, V. W. S. Wong, R. Dobbe, *et al.*, "On Identification of Distribution Grids," *IEEE Trans. Control of Netw. Syst.*, vol. 6, no. 3, pp. 950-960, Sept. 2019.

[31] X. Chen, W. Wu and B. Zhang, "Robust Capacity Assessment of Distributed Generation in Unbalanced Distribution Networks Incorporating ANM Techniques," *IEEE Trans. Sustain. Energy*, vol. 9, no. 2, pp. 651-663, April 2018.

[32] H. Yuan, F. Li, Y. Wei and J. Zhu, "Novel Linearized Power Flow and Linearized OPF Models for Active Distribution Networks With Application in Distribution LMP," *IEEE Trans. Smart Grid*, vol. 9, no. 1, pp. 438-448, Jan. 2018.

[33] E. Dall'Anese, S. S. Guggilam, A. Simonetto, Y. C. Chen and S. V. Dhople, "Optimal Regulation of Virtual Power Plants," *IEEE Trans. Power Syst.*, vol. 33, no. 2, pp. 1868-1881, March 2018.

[34] O. Gandhi, W. Zhang, C. D. Rodríguez-Gallegos, M. Bieri, T. Reindl, and D. Srinivasn, "Analytical approach to reactive power dispatch and energy arbitrage in distribution systems with DERs," *IEEE Trans. Power Syst.*, vol. 33, no. 6, pp. 6522–6533, Nov. 2018.

[35] S.Weckx, C. Gonzalez, and J. Driesen, "Combined central and local active and reactive power control of PV inverters," *IEEE Trans. Sustain. Energy*, vol. 5, no. 3, pp. 776–784, Jul. 2014.

[36] A. Hauswirth, S. Bolognani, G. Hug and F. Dörfler, "Projected gradient descent on Riemannian manifolds with applications to online power system optimization," 2016 54th Annual Allerton Conference on Communication, Control, and Computing (Allerton), Monticello, IL, 2016, pp. 225-232.

[37] R. D. Zimmerman, C. E. Murillo-Sánchez and R. J. Thomas, "MATPOWER: Steady-State Operations, Planning, and Analysis Tools for Power Systems Research and Education," *IEEE Trans. Power Syst.*, vol. 26, no. 1, pp. 12-19, Feb. 2011.

[38] T. Zhao, H. Chiang and K. Koyanagi, "Convergence analysis of implicit Z-bus power flow method for general distribution networks with distributed generators," *IET Gener., Transm. & Distri.*, vol. 10, no. 2, pp. 412-420, 4 2 2016.

[39] S. Bolognani, R. Carli, G. Cavraro and S. Zampieri, "On the Need for Communication for Voltage Regulation of Power Distribution Grids," *IEEE Trans. Control Netw. Syst.*, vol. 6, no. 3, pp. 1111-1123, Sept. 2019.